\documentclass{article}
\usepackage{spconf,amsmath,graphicx}

\usepackage{mathtools}

\usepackage{siunitx}
\usepackage{cite}

\usepackage{hyperref}
\usepackage{cleveref}
\usepackage{amsfonts}
\usepackage{algorithmic}
\usepackage{graphicx}
\usepackage{textcomp}
\usepackage[table,xcdraw]{xcolor}
\usepackage{url}
\usepackage{multirow}
\usepackage{booktabs}
\usepackage{balance}
\usepackage{caption}
\usepackage{balance}
\usepackage{amssymb}
\usepackage{subfig}
\usepackage[flushleft]{threeparttable}

\title{A Sequence Agnostic Multimodal Preprocessing for Clogged Blood Vessel Detection in Alzheimer's Diagnosis}

\name{Partho Ghosh$^{1}$ \quad Md. Abrar Istiak$^{1}$ \quad Mir Sayeed Mohammad$^{1}$ \quad Swapnil Saha$^{1}$ \quad Uday Kamal$^{2}$
}

\address{
$^{1}$ Bangladesh University of Engineering and Technology (BUET), Dhaka-1205, Bangladesh$^{\dagger}$\protect\\
$^{2}$ Georgia Institute of Technology,
Atlanta, GA 30332-0250 USA}

\begin{document}
%
\maketitle
\begin{abstract}

Successful identification of blood vessel blockage is a crucial step for Alzheimer's disease diagnosis. These blocks can be identified from the spatial and time-depth variable Two-Photon Excitation Microscopy (TPEF) images of the brain blood vessels using machine learning methods. In this study, we propose several preprocessing schemes to improve the performance of these methods. Our method includes 3D-point cloud data extraction from image modality and their feature-space fusion to leverage complementary information inherent in different modalities. We also enforce the learned representation to be sequence-order invariant by utilizing bi-direction dataflow. Experimental results on The Clog Loss dataset \footnote[1]{\href{https://www.drivendata.org/competitions/65/clog-loss-alzheimers-research/data/}{https://www.drivendata.org/competitions/65/clog-loss-alzheimers-research/data/}}
show that our proposed method consistently outperforms the state-of-the-art preprocessing methods in stalled and non-stalled vessel classification.  
\end{abstract}
\section{Introduction}
Patients diagnosed with Alzheimer's disease (AD) have been observed to experience cerebral blood flow (CBF) reduction which is the early symptom of this disease \cite{zlokovic2011neurovascular,kisler2017cerebral}. Blocks created by the white blood cells (WBC) adhering to the interior of capillaries are the primary cause of CBF reduction \cite{zlokovic2011neurovascular}. These blocks adversely affect the downstream vessels, and eventually, the overall cerebral blood flow reduces. An experiment \cite{hernandez2019neutrophil} was conducted to check whether the Alzheimer's condition of a mouse would improve by removing the adhesion of the WBC in such stalled capillaries. Experiments showed that the memory function of the mouse improved as blood flow increased \cite{hernandez2019neutrophil}. These results have paved the way for further research to improve the memory functions of an Alzheimer's patient, necessitating the need for more analytical data on stalled capillaries in these studies \cite{hernandez2019neutrophil}. Advanced imaging technologies such as TPEF have enabled researchers to obtain 3D visualizations of extremely narrow blood vessels \cite{haft2019deep,hernandez2019neutrophil}. The crucial step to analyzing such data is the detection and segmentation of these capillaries and labeling them as either stalled or non-stalled in the context of CBF \cite{hernandez2019neutrophil,bracko2020increasing}. However, this requires tremendous time-consuming manual labor that can obscure the research progress. Crowd-scoring method \cite{michelucci2016power} has been proposed as an alternative solution that combines many individual decisions to make a single conclusion. A citizen science application called Stall Catchers \footnote[2]{\href{https://stallcatchers.com/main}{https://stallcatchers.com/main}} 
is such an annotation method. Recent successes of modern deep learning methods in computer vision tasks \cite{litjens2017survey} make them ideal candidates to detect stall and non-stall capillaries with enough accuracy, which can not only reduce the manual labor but also make this research progress faster \cite{hernandez2019neutrophil}.   

\begin{figure*}[htb]
    \centering    \centerline{\includegraphics[width=0.95\linewidth]{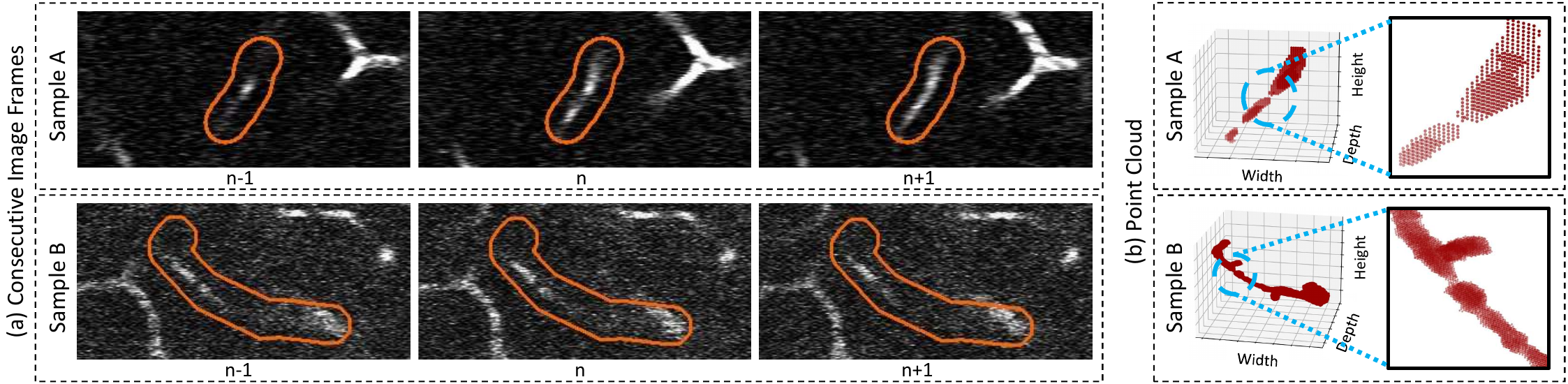}}
    \caption{Visualization of different modalities. Sample A has a better representation of the stalled vessel in the PC while it is not distinct in the image modality. In sample B, the image frames clearly show the stall whereas it is less apparent in the PC.}
    \label{motivation}
\end{figure*}

In the context of binary classification based on the nature of the blood flow, several works have been proposed. Javia \cite{javia2018machine} et. al. have proposed a method for detecting septic/ nonseptic patients by analyzing micro-circulation videos using the t-SNE method. McIlroy \cite{mcilroy2017vivo} et.al. have analyzed videos of animal blood to detect inflammation in blood vessels by using an effective sampling method for training and an average prediction method for prediction. In the case of analyzing TPEF images, deep learning-based methods have been recently used for blood vessel segmentation \cite{haft2019deep,haft2020topological}, cancer cell segmentation and classification \cite{cai2020dense,huttunen2020multiphoton}. In the context of our task and dataset, the state-of-the-art method \cite{solovyev20223d} proposed a 3D convolution network, 3D augmentation, and an ensemble of several networks. 
However, none of these works have used any kind of data-specific preprocessing techniques to tackle the challenges of the dataset, such as eliminating irrelevant pixels, and the sequential variability problem. Recent studies \cite{jeong2018image,helin2022possible} have shown how data-specific preprocessing can improve the deep learning model performance. Inspired by these works, we propose a pipeline of data-specific preprocessing and multimodal fusion techniques to improve the performance of deep learning models in classifying stall and non-stall brain capillaries. Our contributions here are three-fold. First, we introduce a multimodal approach where we use the point cloud (PC) data and image data and process them each in separate streams. The main motivation behind this is visualized in \Cref{motivation}. It is apparent that for some examples it is difficult to identify the stalled capillaries in image data whereas it is quite easy in point cloud data and vice-versa. 
Second, we introduce a background separation scheme to help 
the model better differentiate between the relevant and irrelevant pixels. Third, we propose a bi-direction dataflow technique to learn a sequence-agnostic representation for more condensed attention to the stalled region. 


\section{Methodology}
The key objective is to predict whether a given sample has stalled blood vessels from a video file where the depth of a sample varies with the duration. 
We propose a dual-stream architecture: one that uses the cropped and background (BG) separated video frames as 3D input, and another that uses a point cloud constructed from processing the stacked frames. 

\begin{figure*}[htb]
    \centering    \centerline{\includegraphics[width=\textwidth]{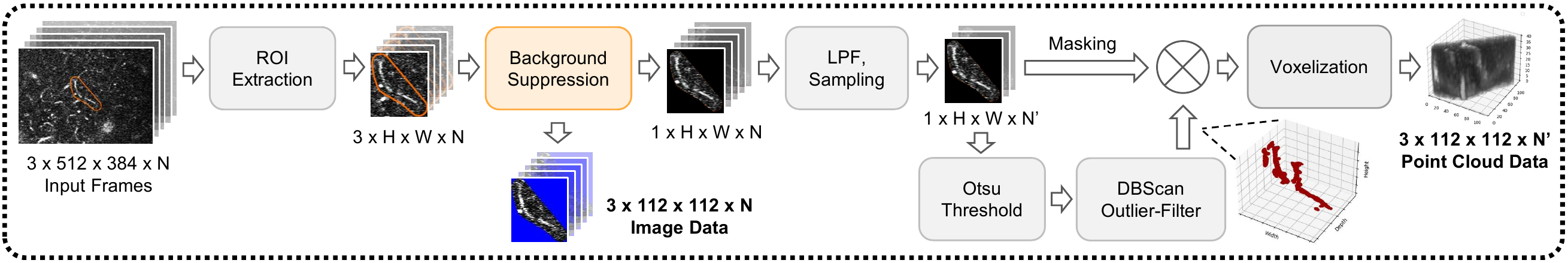}}
    \caption{Data preprocessing steps: input video frames are processed to generate the image and the point cloud data.}
    \label{data_process}
\end{figure*}

\subsection{Data Preprocessing}

\quad\ \textbf{Image Data Preprocessing:} 
For the image-based stream, in \Cref{data_process}, we first extract the region of interest (ROI) from video frames as provided in the dataset. Then we distinguish the region outside the ROI by replacing it with an arbitrary pixel value absent from the original data (blue in our experiments) to help the models focus specifically on the ROI. Finally, the processed image frames are stacked into a 3-D array with 3 color channels $(\mathbb{R}^{3\times112\times112\times N})$.

\textbf{Point Cloud Data Preprocessing:}
For converting the video frames into a volumetric representation of blood vessels (\Cref{data_process}), we first extract the region of interest from the image frames and stack them depth-wise to get a 3D volume. Then we apply a 3D Gaussian low pass filter (LPF) to the image stack and sample frames to reduce the dimension. We apply otsu \cite{otsu1979threshold} threshold to the 3D volume to create a binary mask, which gives a point cloud representation of the blood vessels. We use DBSCan clustering\cite{ester1996density} to the point cloud mask that removes the outliers and gives a noise-free 3D mask. Finally, in the voxelization step, this mask is applied to the 3D stacked image frames for suppressing non-vessel regions from the 3D volume $(\mathbb{R}^{3\times112\times112\times N'})$. 

\subsection{Dual Stream Network with Multimodal Fusion}

\begin{figure}[t]
    \centering
    \includegraphics[width = \linewidth]{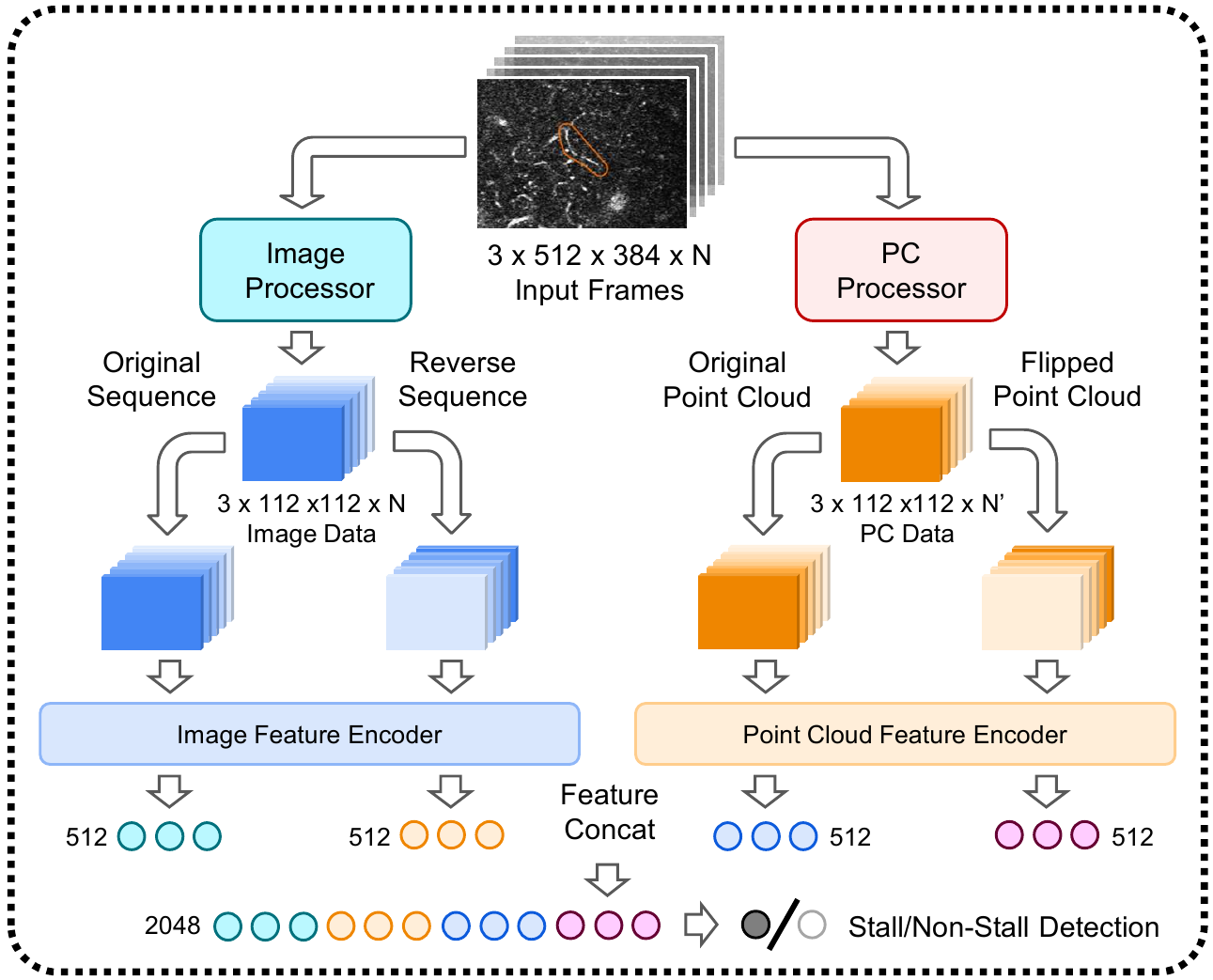} 
    \caption{Proposed Multimodal Model Fusion Architecture.}
    \label{archi}
\end{figure}

In proposed approach represented in \Cref{archi}, each of the two data streams consists of a feature encoder \emph{E} that takes a 3D volume 
$\mathbf{X} = \{x_i|x \in \mathbb{R}^{3\times 112 \times 112}, 1 \leq i \leq n\} $
 as input ($x_i$ = $i$-th frame, $n$ = no. of frames) and outputs a feature vector $\mathbf{F}$ $\in  \mathbb{R}^{1\times 512}$. For the image stream, we pass the image stack through the encoder network in a bi-directional manner and concatenate the two generated feature vectors into $\mathbf{\hat{F_I}}$ $\in  \mathbb{R}^{1\times 1024}$:
\begin{equation} \label{bi_direc_img}
      \mathbf{\hat{F_I}} = \{E_I([x_1, x_2, ...., x_n]), E_I([x_n, x_{n-1}, ...., x_1])\}  
\end{equation}
Similarly, we pass the point cloud data bidirectionally through another feature encoder, creating a second concatenated feature vector $\mathbf{\hat{F_P}}\in  \mathbb{R}^{1\times 1024}$. A final dense layer $M$ followed by the sigmoid activation function ($\sigma$)  generates the binary prediction score from the 
fusion of image and point cloud feature vectors:
\begin{equation} \label{final_pred} 
       score = \sigma(M(\{\mathbf{\hat{F_I}}, \mathbf{\hat{F_P}}\}))
\end{equation}


\section{Experiments and Results}
In this section, we discuss the experimental setup for stalled vs non-stalled classification in detail. We also study the effectiveness of each component of our method from the systematic ablation experiments described here. 

\textbf{Dataset:}\label{dataset}
The Clog Loss dataset for Alzheimer's research contains stacks of 3D volumetric images saved in video format where the depth of the volume varies with the video time frame. The dataset is collected from the TPEF microscopy of live mouse brains showing the flow inside brain blood vessels that appear as bright interconnected regions. The videos have a resolution of $512 \times 384$ pixels and a variable duration. Regions of interest on the frames are marked with a orange overlay. The main objective of the dataset is to detect stalled vessels inside a TPEF volumetric sample, i.e. classification of the samples based on the presence of a discontinuity in the white vessel structures that characterize a clogged blood vessel. To evaluate the effectiveness of our proposed dual-stream network, we use the `Nano subset' of the clog loss data containing 1413 video samples with 50\% stalled and 50\% of non-stalled instances. For training, validation, and testing purposes the nano subset was split into a 75:15:10 ratio.

\textbf{Experimentation Setup:}
We perform the experiments using PyTorch on a system with a Tesla P100 16 GB GPU. We used a batch size of 1, Adam optimizer with an initial learning rate $10^{-4}$, binary cross-entropy with logits loss for the classification loss function, and accuracy and Matthews Correlation Coefficient (MCC)\footnote[4] {\href{https://en.wikipedia.org/wiki/Phi\_coefficient}{https://en.wikipedia.org/wiki/Phi\_coefficient}} for the evaluation metric. 

\subsection{Ablation Studies}\label{ablation_section}

\begin{figure*}[htb]
    \centering    \centerline{\includegraphics[width=0.95\textwidth]{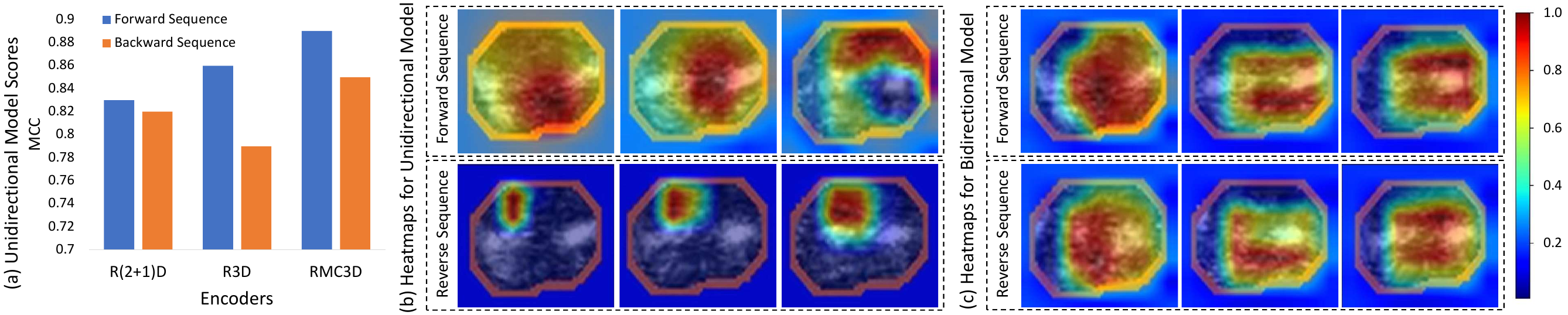}}
    \caption{(a) Score decreases when the input to the unidirectional model is reversed and (b) it fails to identify stalls as shown in the 2nd row. (c) The bidirectional model attends consistently to the stalled region for both sequences.}
    \label{dataflow_vis}
\end{figure*}

\quad\ \textbf{Feature Encoder:}
We evaluate the performance with various state-of-the-art 3D classification networks, namely - ResNet(2+1)D, R3D, and MC3D, (which are readily available from Torchvision library\footnote[5]{\footnotesize \href{https://pytorch.org/vision/stable/models.html}{https://pytorch.org/vision/stable/models.html}}) as our backbone feature extractor to study the generalization of our proposed multi-modal fusion technique. Initially, we train each of the three encoders with the processed point cloud data and background-separated image data individually. All these experiments show a consistent performance improvement with the multi-modal fusion approach for the different encoders, as shown in  \Cref{compared_models}. It is apparent that the R(2+1)D network performs slightly better (MCC) when using multimodal data and thus, we use this encoder for the subsequent experiments.

\begin{table}[ht]
\caption{Performance Comparison on Different Encoders}
\label{compared_models}
\resizebox{\columnwidth}{!}{\begin{tabular}{ccccccc}
\midrule
\multicolumn{1}{c}{\multirow{3}{*}{Data*}} & \multicolumn{6}{c}{Feature Encoder}                                                                                                                                 \\ \cmidrule{2-7} 
\multicolumn{1}{c}{}                                    & \multicolumn{2}{c}{R3D}                                  & \multicolumn{2}{c}{MC3D}                                 & \multicolumn{2}{c}{R(2+1)D}         \\ \cmidrule{2-7} 
\multicolumn{1}{c}{}                                    & \multicolumn{1}{c}{Acc}    & \multicolumn{1}{c}{MCC}    & \multicolumn{1}{c}{Acc}    & \multicolumn{1}{c}{MCC}    & \multicolumn{1}{c}{Acc}    & \multicolumn{1}{c}{MCC}    \\ \midrule
PC                                          & \multicolumn{1}{l}{0.9167} & \multicolumn{1}{l}{0.8312} & \multicolumn{1}{l}{0.8819} & \multicolumn{1}{l}{0.7629} & \multicolumn{1}{l}{0.8611} & 0.7237 \\ \midrule
BGS                           & \multicolumn{1}{l}{0.9296} & \multicolumn{1}{l}{0.8594} & \multicolumn{1}{l}{0.9235} & \multicolumn{1}{l}{0.8472} & \multicolumn{1}{l}{0.9155} & 0.8312 \\ \midrule
\textbf{MM}                                          & \multicolumn{1}{l}{\textbf{0.9366}} & \multicolumn{1}{l}{\textbf{0.8732}} & \multicolumn{1}{l}{\textbf{0.9366}} & \multicolumn{1}{l}{\textbf{0.8732}} & \multicolumn{1}{l}{\textbf{0.9366}} & \textbf{0.8736} \\ \bottomrule
\end{tabular}}
\begin{tablenotes}
      \footnotesize
      \item * PC: Point Cloud, BGS: Background Separated Image, MM: MultiModal
    \end{tablenotes}
\end{table}

\textbf{Background Separation (BGS):}
From the 1st and 2nd rows of \Cref{ablation}, we can see that for the image data, separating the background by replacing it with blue pixels (\Cref{data_process}) increases both the accuracy and MCC by $\sim 1\%$.

\textbf{Bi-directional Dataflow (Bi-DF):}
After thoroughly experimenting with the dataset, we found that the network tries to classify the data by looking at the sequence as it is presented. \Cref{dataflow_vis} (a) shows a notable performance drop when we feed the data in reverse order to an encoder trained using the original sequence(Unidirectional) only. We also visualize the attention heatmap for this experiment in \Cref{dataflow_vis} (b) where it is apparent that the learned representation is sequence order dependent. Introducing bi-directional dataflow helps the model to overcome this as shown in \Cref{dataflow_vis} (c). This is also proved experimentally in rows 2, 3 (for image data) and rows 4, 5 (for point-cloud data) in \Cref{ablation} by an improvement of $4\%$ in accuracy and more than $7\%$ in MCC score.

\textbf{Threshold (Th$^{a/b}$):}
We verify the effectiveness of the data-adaptive threshold (Otsu's method) for generating the point cloud data by comparing it with a fixed-value threshold. We generate the point cloud data by thresholding the image stacks with 90, 95, and 99 percentile pixel values for the three channels that provided the best visual appearance of the vessel structures. We can see from rows 4 and 5 of \Cref{ablation} that the data-adaptive threshold significantly improves the overall performance.

\textbf{Data Modalities:}
In \Cref{ablation}, the performance gain after adding each component is summarized. For independent modality, incorporating all our proposed components result in a $4$\% increase in accuracy, $10$\% increase in MCC for the image data, and a $6$\% improvement in accuracy, $12$\% improvement in MCC for the point cloud data. Finally, the multimodal fusion in the last row shows the best result among all the previous experiments.

\begin{table}[t]
\caption{Ablation Study (on R(2+1)D Encoder). The ID column denotes the use of the image data and the PC column denotes the use of point cloud data.}
\label{ablation}
\centering
\resizebox{3.4in}{0.8in}{
\begin{tabular}{ccccccc}
\toprule
ID & PC & BGS & Bi-DF & Th$^{a/b}$ & Acc & MCC \\ \midrule
\checkmark & - & - & - & - & 0.9085 & 0.8191\\
\checkmark & - & \checkmark & - & - & 0.9155 & 0.8312 \\
\checkmark & - & \checkmark & \checkmark & - & 0.9437 & 0.8883 \\
\midrule
- & \checkmark & - & - & a & 0.8611 & 0.7237 \\
- & \checkmark & - & - & b & 0.8819 & 0.7629 \\
- & \checkmark & - & \checkmark & b & 0.9225 & 0.8453 \\
\midrule
\pmb{\checkmark} & \pmb{\checkmark} & \pmb{\checkmark} & \pmb{\checkmark} & \textbf{b} & \textbf{0.9507} & \textbf{0.9019} \\
\bottomrule
\end{tabular}
}
\begin{tablenotes}
      \footnotesize
      \item ID: image data, PC: point cloud data, BGS: background separation, Bi-DS: bi-directional dataflow,  Th$^a$: percentile threshold, Th$^b$: Otsu threshold.
    \end{tablenotes}
\addtolength{\tabcolsep}{10pt}
\end{table}

\textbf{Comparison with SoTA Preprocessing:}
The performance benchmark of the dataset that we used in this work has a private leaderboard. We compare the performance of our preprocessing method with that of the current top-scoring solutions. While none of these methods utilized any data or task-specific preprocessing (like we proposed), they leveraged different transformation techniques on the raw data during training before passing them to the network. The leading solution\cite{solovyev20223d} utilizes different processing methods including spatial, pixel, and custom-made transformations. The 2nd placeholder adopts a larger frame size (160x160) and spatial transformations including horizontal and vertical flips, rotation, distortions, and random gaussian noise addition as their preprocessing. Finally, the third placeholder's preprocessing consists of flip and rotation. In \Cref{sota-table} performances of the discussed methods as well as ours are shown. For a fair comparison, we use the R(2+1)D encoder for all the experiments. Surprisingly, the 1st and 2nd placeholders preprocessing methods did not perform as expected on the R(2+1)D encoder. We hypothesize the reason can be their preprocessing methods are not generalizable across different encoders. Nonetheless, from \Cref{sota-table} we can see that our proposed method surpasses the best result by more than $4\%$ in accuracy and $8\%$ in MCC.

\begin{table}[h!]
\caption{Performance comparison of R(2+1) Encoder with top scoring methods' preprocessing.}
\label{sota-table}
\centering
\begin{tabular}{ccc}
\toprule
Preprocessing & Acc & MCC    \\ \midrule
1st Place                       & 0.7958                   & 0.5978 \\ 
2nd Place                 & 0.8380                    & 0.6801 \\ 
3rd Place           & 0.9085                   & 0.8191 \\ \midrule
\textbf{Proposed}                     & \textbf{0.9507}                   & \textbf{0.9019} \\ 
\bottomrule
\end{tabular}

\begin{tablenotes}
      \footnotesize
      \item \href{https://github.com/drivendataorg/clog-loss-alzheimers-research}{https://github.com/drivendataorg/clog-loss-alzheimers-research}
\end{tablenotes}

\end{table}


\section{Discussion \& Conclusion}

In this work, we propose a multimodal fusion-based data-specific preprocessing pipeline to improve the classification performance of stalled vs non-stalled brain capillaries. Our extensive experiments on the Clog Loss dataset show the efficacy and significance of our proposed method which consistently outperforms the existing solutions. One of the limitations of this work is that we considered off-the-shelf feature encoders which may not be optimal for the task at hand. Future work can be in the direction of designing task-specific neural architecture that will leverage the unique properties of the proposed multi-modal, bi-directional data streams, which we hope, will significantly improve the research progress in Alzheimer's disease diagnosis.

\bibliographystyle{IEEEbib}
\bibliography{refs}

\end{document}